\documentclass[sigconf]{acmart}
\usepackage{algorithm}
\usepackage{algpseudocode}
\usepackage{booktabs}
\usepackage{tabularx}
\usepackage{listings}
\usepackage{xcolor} 
\usepackage{amsmath}
\usepackage{pifont} 
\usepackage{breqn}

\begin{document}
\pagestyle{plain} 
\settopmatter{printacmref=false} 
\renewcommand\footnotetextcopyrightpermission[1]{} 

\title{An Efficient Algorithm for Generating Minimal Unique-Cause MC/DC Test Cases for Singular Boolean Expressions}


\author{Robin Lee}
\email{binroi@gnu.ac.kr}
\orcid{0000-0003-3871-0924}
\affiliation{%
  \institution{Department of Computer Science and Engineering, Gyeongsang National University}
  \city{Jinju}
  \country{Republic of Korea}
}

\author{Youngho Nam}
\authornote{Corresponding author}
\email{yhnam@gnu.ac.kr}
\orcid{0009-0001-7482-8340}
\affiliation{%
  \institution{Department of Computer Science and Engineering, Gyeongsang National University}
  \city{Jinju}
  \country{Republic of Korea}
}

\renewcommand{\shortauthors}{et al.}

\begin{abstract}
Modified Condition/Decision Coverage (MC/DC) is a mandatory structural coverage criterion for assuring the reliability of safety-critical software. Among its variants, Unique-Cause MC/DC provides the strongest assurance, yet efficient and scalable test generation for Unique-Cause MC/DC remains underexplored. This gap is particularly important because large-scale avionics studies report that 99.7\% of conditional decisions are Singular Boolean Expressions (SBEs), for which Unique-Cause obligations can be precisely characterized. We propose \emph{Robin’s Rule}, a deterministic, direct-construction algorithm that generates a provably minimal test suite of $N+1$ test cases to guarantee 100\% Unique-Cause MC/DC for SBEs with $N$ conditions, without enumerating the $2^N$ truth table. The algorithm runs in $O(N^2)$ time by explicitly constructing an $(N+1)\times N$ test table. To evaluate the approach, we build a benchmark of 25 SBEs consisting of 15 TCAS-II–derived conditions and 10 randomly generated SBEs with diverse operators and nesting. We validate achieved coverage using VectorCAST as an oracle and compare against state-of-the-art baseline paradigms using BDD and SAT. Across all benchmarks, Robin’s Rule consistently achieves 100\% Unique-Cause MC/DC with the theoretical minimum number of tests, while providing stable and efficient generation time. This work offers a practical and provably optimal solution for Unique-Cause MC/DC test generation on SBEs, improving both rigor and scalability for safety-critical verification.
\end{abstract}




\maketitle



\section{Introduction}

Modern society relies heavily on complex and sophisticated software systems across various domains, including avionics, automotive, medical devices, and industrial control, where a failure can lead to catastrophic human and material losses, making reliability and safety a paramount concern~\cite{Martins2020}, \cite{An2021}, \cite{Singh2021}, \cite{Singh2018}. Within the software development lifecycle, the Verification and Validation (V\&V) activities play a critical role in ensuring this reliability, with software testing standing as the primary method for identifying and eliminating potential defects before deployment. However, as software complexity increases, the time and cost required for thorough testing can become prohibitive, often consuming over 50\% of a project's budget and requiring repetitive execution for every code change.

To manage these costs while quantitatively evaluating the effectiveness of testing, the industry widely employs code coverage metrics. Code coverage serves as a measure of test adequacy by indicating the degree to which source code has been exercised by a particular test suite. While criteria like Statement, Branch, and Condition Coverage exist, they often fail to detect intricate logical faults within complex conditional expressions. At the other end of the spectrum, Multiple Condition Coverage (MCC) is the most thorough, requiring $2^N$ test cases for $N$ conditions to cover every possible logical combination, but its combinatorial explosion renders it impractical for all but the simplest cases.

Against this backdrop, \textbf{Modified Condition/Decision Coverage (MC/DC)} was established as an effective and practical compromise \cite{Squire2020}. The core principle of MC/DC is to demonstrate that each individual condition can independently affect the decision's outcome, regardless of the other conditions. Due to this blend of rigor and efficiency, MC/DC is mandated by numerous international safety standards for software at the highest integrity levels, including DO-178C (avionics) \cite{DO178C}, ISO 26262 (automotive) \cite{ISO26262}, IEC 61508 (industrial) \cite{IEC61508}, and IEC 62304 (medical) \cite{IEC62304}, EN 50128 (railway) \cite{EN50128}.

MC/DC is primarily categorized into \textit{Unique-Cause} and \textit{Masking} forms \cite{Chilenski2001}. \textit{Unique-Cause} MC/DC is the stricter approach that most clearly guarantees a condition's independence by requiring that only the condition of interest changes its value between a pair of tests. In contrast, \textit{Masking} MC/DC offers more flexibility by allowing for logical masking like short-circuit evaluation \cite{CAST2001}. However, this flexibility can lead to subtle issues. For instance, research by Rajan et al.\cite{Rajan2014} points out that developers can intentionally alter code structure (e.g., through inlining) to more easily satisfy coverage requirements, a practice sometimes referred to as \textbf{`cheating'}. This possibility, arising from the flexibility of \textit{Masking} MC/DC, underscores the need for research into the more robust \textit{Unique-Cause} MC/DC for systems that demand the highest level of reliability.

Despite its importance, a significant portion of existing research has focused on \textit{Masking} MC/DC. This trend was largely driven by the "solvability problem," as the strict rules of \textit{Unique-Cause} MC/DC make it impossible to achieve coverage for many real-world expressions containing coupled conditions (Non-SBEs), where the same condition variable appears more than once. However, this problem is rare in practice. An analysis of large-scale avionics software revealed that \textbf{99.7\% of all conditional decisions were, in fact, Singular Boolean Expressions (SBEs)}, with only 70 out of 20,256 being Non-SBEs \cite{Chilenski2001}. An SBE, an expression where each condition variable appears only once, is the ideal structure for applying \textit{Unique-Cause} MC/DC to unambiguously prove the independence of each condition. This implies that a highly efficient test generation technique optimized for SBEs would be both practical and highly impactful for the vast majority of real-world safety-critical systems.

Therefore, this paper proposes a new algorithm, \textbf{`Robin's Rule'}, to overcome the limitations of existing test generation methods, such as the exponential complexity of exploring a full truth table ($2^N$). Our algorithm directly generates the theoretical minimum of $N+1$ test cases for SBEs, aiming to dramatically reduce verification costs while satisfying the strictest testing standards.

The contributions of this paper are threefold and are detailed as follows:
\begin{enumerate}
    \item \textbf{A Deterministic Algorithm for Minimal Test Generation for Unique-Cause MC/DC:} We introduce \textbf{`Robin's Rule'}, a deterministic and procedural algorithm that guarantees the theoretical minimum $N+1$ test set for \textit{Unique-Cause} MC/DC on SBEs. This algorithm overcomes the limitations of existing complex search or constraint-solving methods by analyzing the structural properties of the expression and directly constructing the test cases according to explicit rules.
    \item \textbf{Experimental Demonstration of Correctness and Optimality:} We empirically evaluate the correctness and optimality of the proposed algorithm. Through extensive experiments on a benchmark dataset based on the real-world safety-critical TCAS-II specification, we show that the generated $N+1$ test cases consistently achieve 100\% Unique-Cause MC/DC coverage across all SBEs when validated using VectorCAST as an independent coverage oracle.    
    \item \textbf{A Practical Verification Solution for Safety-Critical Systems:} We present a practical and cost-effective solution for the verification of safety-critical systems. The results of this study provide a concrete method to dramatically reduce significant testing overhead and costs by reducing the number of test cases to the theoretical minimum, without compromising the strictest reliability standards.
\end{enumerate}

The remainder of this paper is organized as follows. Section 2 discusses background and related work. Section 3 details the proposed algorithm. Section 4 presents the experimental evaluation and discusses the results. Finally, Section 5 concludes the paper and outlines future work.


\section{Background and Related Work}


\subsection{Preliminaries}


\subsubsection{The Need for and Definition of MC/DC}
\leavevmode\par
In software testing, code coverage is a critical metric for test adequacy. While Decision Coverage can miss logical errors within individual conditions, Multiple Condition Coverage (MCC) is often infeasible due to the exponential ($2^N$) growth of test cases. MC/DC (Modified Condition/Decision Coverage) was proposed as a practical compromise, requiring that each condition be shown to independently affect the decision's outcome \cite{Hayhurst2001}, \cite{Chilenski2001}. This independence is verified by finding an `independence pair' for each condition: a pair of test cases where only the value of that single condition is toggled, which in turn toggles the final outcome of the overall decision.

This notion of independent effect can be analyzed formally using the Boolean Difference Function, as shown in Equation~\eqref{eq:boolean_difference}.
\begin{equation}\label{eq:boolean_difference}
    \frac{\delta F(c)}{\delta c_i} = F(c_1, \dots, c_i, \dots, c_n) \oplus F(c_1, \dots, \neg c_i, \dots, c_n)
\end{equation}
where $\oplus$ denotes the XOR operation. A result of `True' for this function signifies that toggling the value of $c_i$ while all other conditions are held constant will toggle the outcome of the entire function $F(c)$. This function, therefore, serves as the mathematical basis for determining if a condition has an independent influence for a given input vector $c$.


\subsubsection{Formal Definitions of Unique-Cause and Masking MC/DC}
\leavevmode\par
Using the Boolean Difference Function, the two primary forms of MC/DC can be defined more rigorously \cite{Chilenski2001}, \cite{CAST2002}.

\paragraph{Unique-Cause MC/DC} This is the strictest form, requiring that an independence pair, consisting of two test vectors $x$ and $y$ for a condition $c_i$, must satisfy all of the following rules \cite{Hayhurst2001}:
\begin{enumerate}
    \item $x_i \neq y_i$
    \item $F(x) \neq F(y)$
    \item For all $j \neq i$, $x_j = y_j$
\end{enumerate}
This third rule, "holding all other conditions fixed," is the core constraint that ensures the change in condition $c_i$ is the \textbf{`unique cause'} for the change in the outcome.

\paragraph{Masking MC/DC} This is a more flexible form that requires an independence pair $x$ and $y$ for a condition $c_i$ to satisfy the following \cite{CAST2001}:
\begin{enumerate}
    \item $x_i \neq y_i$
    \item $F(x) \neq F(y)$
    \item $\frac{\delta F(x)}{\delta x_i} = \text{True} \land \frac{\delta F(y)}{\delta y_i} = \text{True}$
\end{enumerate}

\subsubsection{Singular Boolean Expressions (SBE) and Coupled Conditions}
\leavevmode\par
An SBE is defined as a non-degenerate Boolean expression where each condition variable appears only once. In contrast, a Non-SBE such as `(A \&\& B) || (A \&\& C)' contains Coupled Conditions. Proving the independence of the first instance of `A' while holding the second instance of `A' fixed is logically impossible, making Unique-Cause MC/DC inapplicable. The fact that 99.7\% of conditions in large-scale avionics systems are SBEs reinforces the practical importance of this study \cite{Chilenski2001}.


\subsection{Related Work}

The automated generation of test cases for MC/DC is a well-established research area that has been actively studied over the past decades \cite{Cao2024}, \cite{Shekhawat2021}, \cite{Barisal2021}, \cite{Godboley2017}, \cite{Kangoye2015}, \cite{Haque2014}, \cite{Ghani2009}, \cite{Awedikian2009}. Early and theoretical approaches included using the Boolean Difference Function or the Independence Graph model to theoretically prove the feasibility of constructing a minimal $N+1$ test set. Modern research has built upon this foundation, proposing more sophisticated and automated generation techniques.


\subsubsection{Formal Model and Constraint Solving-Based Generation}
\leavevmode\par
This family of approaches transforms the MC/DC problem into a formal model to be solved by powerful engines. The most active research area involves leveraging \textbf{SAT/SMT solvers}. This technique formalizes all MC/DC rules (e.g., independence pairs) as logical constraints and uses a solver to find a minimal set of test inputs that satisfies them. The works of Kitamura et al.\cite{Kitamura2018}, Yang et al.\cite{Yang2018}, and Jaffar et al.\cite{Jaffar2019} are representative of this category. While powerful, they often target \textit{Masking} MC/DC and their internal search process can be a `black-box'. Furthermore, \textbf{model checkers} like CBMC have been utilized to explore the program's state space to find paths satisfying MC/DC \cite{Golla2024}.
Reinforcing this trend, the work by Golla et al.\cite{Golla2025} specifically targets the stricter \textbf{Unique-Cause MC/DC}. Their proposed `UCM-Gen' tool generates all possible inequality comparison sequences for a predicate, which are then annotated into the program. The CBMC model checker is subsequently used to formally verify these sequences and compute the UCM Score. This approach highlights a key distinction: while effective, it relies on a powerful, general-purpose verification engine to solve the UCM problem, contrasting with our direct, deterministic algorithm.

Another prominent technique involves transforming Boolean expressions into graph-based data structures. \textbf{Binary Decision Diagrams (BDDs)} represent a Boolean function in a compressed, directed acyclic graph, avoiding the exponential space problem of a full truth table. Test case generation is performed by searching for specific paths on this graph to find combinations that satisfy the independence pairs. However, BDDs are sensitive to variable ordering, and pathfinding often relies on heuristics. The research by Ahishakiye et al.~\cite{Ahishakiye2021} demonstrated success by combining roBDDs with heuristic pathfinding to generate the $N+1$ minimal set for a significant portion of the TCAS-II benchmark, but did not guarantee minimality for all SBEs.


\subsubsection{Search-Based and Heuristic Approaches}
\leavevmode\par
These methods frame MC/DC coverage attainment as an optimization problem and employ various search techniques. \textbf{Search-based heuristics}, such as genetic algorithms \cite{Kaur2011} or reinforcement learning \cite{Cegin2020}, are effective in complex search spaces but are often non-deterministic and do not formally guarantee finding the theoretical optimum. Recent work by Sartaj et al. \cite{Sartaj2025} has extended this approach into the Model-Based Testing (MBT) domain. They focus on generating MC/DC test data from high-level system constraints written in OCL (Object Constraint Language). Their search-based approach incorporates heuristics like Case-Based Reasoning (CBR) and Range Reduction to improve efficiency. While this presents a practical approach for testing at a higher level of abstraction, it remains fundamentally reliant on search-based heuristics, distinguishing it from our deterministic method.


\subsubsection{Code Transformation-Based Approaches}
\leavevmode\par
Another interesting line of research involves \textbf{Code Transformation} techniques to facilitate test generation. The `SMUP' technique proposed by Barisal et al.\cite{Barisal2024} is a representative example. This method modifies the source code by inserting additional `if-else' blocks with specific patterns around the original conditional statements. The transformed code, while semantically equivalent to the original, is structured to provide new execution paths that guide a concolic testing tool, like jCUTE, to more easily discover MC/DC independence pairs. This work focuses on `improving' MC/DC coverage on existing code, which is a fundamentally different philosophy from our approach of directly generating a minimal set from an unmodified SBE.


\subsubsection{Research Trends and Positioning of This Work}
\leavevmode\par
There is an important reason why Masking MC/DC became the mainstream approach among these diverse studies. The strict "hold all other conditions fixed" rule of \textit{Unique-Cause} MC/DC makes it impossible to achieve coverage for many real-world expressions, especially Non-SBEs that contain \textbf{coupled conditions} \cite{Chilenski2001}. Due to this "unsolvable" problem, many researchers and tool developers adopted the more pragmatic \textit{Masking} MC/DC as a practical alternative with wider applicability \cite{CAST2001}. However, this focus on pragmatism has, paradoxically, resulted in \textbf{slowing down research into optimized solutions for SBEs}, which constitute 99.7\% of actual avionics system code.

Therefore, Robin's Rule possesses distinct differentiation and originality. Unlike the approaches above that rely on complex external technologies (OCL, SMT, BDDs), heuristics, or code transformation, it provides a deterministic and transparent procedure based on the structural rules of the expression, for the clear scope of SBEs, to \textbf{directly construct the $N+1$ minimal set from the start without generating a full truth table}.


\section{The Proposed Algorithm: Robin’s Rule}\label{sec:algorithm}

\begin{figure}[h!]
    \centering
    \includegraphics[width=0.8\columnwidth]{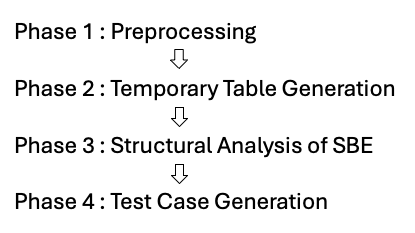}
    \caption{The four-phase process of Robin's Rule.}
    \label{fig:process_flow}
\end{figure}

The core philosophy of \textbf{Robin's Rule} is to bypass the inefficient process of generating a full truth table ($2^N$) and then searching for independence pairs. Our algorithm overcomes this limitation by directly constructing the theoretical minimum of $N+1$ test cases from the start, by analyzing the structural properties and operator relationships of the SBE. This deterministic, multi-phase algorithm is transparent and predictable, without relying on complex external technologies or heuristics. The overall process consists of four main phases as shown in \textbf{Figure~\ref{fig:process_flow}}.

\subsection{Phase 1: Preprocessing}
The primary goal of this phase is to transform any given SBE into a standardized format to ensure consistent processing. This provides a foundation for the algorithm to apply its rules uniformly. The process is detailed in \textbf{Algorithm~\ref{alg:al1}}.

\begin{algorithm}
\caption{Preprocessing of the SBE}\label{alg:al1}
\begin{algorithmic}[1]
\State \textbf{Input:} An SBE, $S$
\State \textbf{Output:} A rearranged SBE, $R$
\Procedure{Pre-Alignment-Process}{$S$}
    \If{$S$ has a top-level NOT operator}
        \State $R \gets \text{apply\_DeMorgan\_recursively}(S)$ \Comment{Phase 1.1: Apply De Morgan's}
        \State $R \gets \text{sort\_conditions}(R)$ \Comment{Proceed to Phase 1.2}
    \Else
        \State $R \gets \text{sort\_conditions}(S)$ \Comment{Phase 1.2: Sort directly}
    \EndIf
    \State \textbf{return} $R$
\EndProcedure
\Statex
\Function{apply\_DeMorgan\_recursively}{expression}
    \If{expression contains NOT}
        \State expression $\gets$ deMorgan(expression) \Comment{Apply to top-level NOT}
        \State \textit{// Recursively process sub-expressions}
        \State expression $\gets$ [apply\_DeMorgan\_recursively(sub\_expr) for sub\_expr in expression]
    \EndIf
    \State \textbf{return} expression
\EndFunction
\Statex
\Function{sort\_conditions}{expression}
    \State \textit{// Sort the main conditions by operand count (descending order)}
    \State expression $\gets$ sort\_by\_operand\_count(expression)
    \State \textit{// Sort sub-conditions within each main condition}
    \State expression $\gets$ [sort\_conditions(sub\_expr) for sub\_expr in expression]
    \State \textbf{return} expression
\EndFunction
\end{algorithmic}
\end{algorithm}

\begin{enumerate}
    \item \textbf{Normalization of NOT Operations}: The algorithm recursively applies De Morgan’s Laws to the SBE. For example, an expression `!(A \&\& B)' is transformed into `!A || !B'. This process ensures that all `NOT' operators apply only to individual conditions (operands), which simplifies subsequent operator analysis.
    \item \textbf{Structural Sorting}: The SBE is then sorted. The main logical blocks connected by the top-level operators are arranged in descending order based on their complexity (i.e., the number of operands they contain). This sorting is applied recursively within each sub-expression. This establishes a deterministic processing order, ensuring that more logically complex parts are handled first.
\end{enumerate}


\subsection{Phase 2: Temporary Table Generation}
Based on the preprocessed SBE, we initialize an empty table to store the final test cases. For an SBE with $N$ operands, we allocate a two-dimensional table of size $(N+1)\times N$, where each row corresponds to a single test case and each column corresponds to a condition. \textbf{Algorithm~\ref{alg:al2}} shows that the table is allocated with exactly $N+1$ rows (or $3$ rows when $N=2$, i.e., $N+1$), and subsequent phases only populate entries without changing the number of rows. Thus, Robin's Rule outputs exactly $N+1$ test cases. It distinguishes between cases where the number of conditions is two versus more than two. If there are only two conditions, it proceeds directly to test case generation based on the operator (AND/OR).

\begin{algorithm}
\caption{Pseudocode for Generating a Temporary Table}\label{alg:al2}
\begin{algorithmic}[1]
\State \textbf{Input:} A rearranged SBE, $R$
\State \textbf{Output:} A temporary table for storing test cases
\Procedure{Temporary\_Table\_Generation\_Process}{$R$}
    \State Let $N$ be the number of operands in $R$
    \If{$N > 2$}
        \State $rows \gets N+1$
        \State $columns \gets N$
        \State temporary\_table $\gets$ generate\_table($rows, columns$)
        \State Call\_Next\_Phases(temporary\_table) \Comment{Proceed to Phase 3 \& 4}
    \ElsIf{$N = 2$}
        \State $rows \gets 3$
        \State $columns \gets 2$
        \State temporary\_table $\gets$ generate\_table($rows, columns$)
        \State Let $B$ be the operator in $R$
        \If{$B$ is AND}
            \State Generate\_Test\_Cases(temporary\_table, "AND")
        \ElsIf{$B$ is OR}
            \State Generate\_Test\_Cases(temporary\_table, "OR")
        \EndIf
    \Else
        \State \textbf{return} Size\_Error
    \EndIf
\EndProcedure
\end{algorithmic}
\end{algorithm}


\textbf{Lemma 1 (Test-suite cardinality).}
For any input SBE with $N\ge 2$ operands, Robin's Rule generates exactly $N+1$ test cases.

\textit{Proof.}
In Phase~2, Algorithm~\ref{alg:al2} allocates a temporary table with \texttt{rows}$\leftarrow N+1$ and \texttt{columns}$\leftarrow N$ when $N>2$. When $N=2$, Algorithm~\ref{alg:al2} sets \texttt{rows}$\leftarrow 3$, which equals $N+1$.
In Phase~3 and Phase~4, the algorithm only fills entries in the pre-allocated table and does not introduce additional rows.
Therefore, the number of output test cases (one per row) is exactly $N+1$ for all $N\ge 2$.
\hfill $\square$


\subsection{Phase 3: Structural Analysis of SBE}

In this phase, the structure of the normalized SBE is analyzed to establish the rules and blueprint for deterministically assigning truth values in Phase 4. To achieve this, this section first defines the core terminology and notation used by the algorithm, and then describes the analysis process, which involves decomposing the SBE into structural forms and then generating a relation table.

\subsubsection{Terminology and Notation}
\leavevmode\par
To systematically analyze any given SBE, Robin's Rule uses a set of standardized terms and symbols. These are defined in \textbf{Table~\ref{tab:notation}}. This table serves as a glossary for understanding the components of the algorithm described hereafter.

\begin{table}[h!]
  \centering
  \caption{Terminology and Notation for SBE Analysis}
  \label{tab:notation}
  \begin{tabularx}{\columnwidth}{c l X}
    \toprule
    \textbf{Symbol} & \textbf{Term} & \textbf{Description} \\
    \midrule
    $Opr_i$ & Operand & An individual condition variable (e.g., \texttt{A}, \texttt{B}, \texttt{!C}). \\
    $i$ & Index & A sequence number for operands, B-forms, or S-forms, where $i \in \{1, \dots, N\}$. \\
    $o$ & Generalized Operator & A symbol for a logical AND ($\wedge$) or OR ($\vee$) operator. \\
    $S_i$ & S-Form & The most basic logical unit, consisting of a single operand ($Opr_i$). \\
    $B_i^o$ & B-Form & A two-operand logical block where two operands are connected by one operator ($B_i^o = Opr_i \circ Opr_{i+1}$). \\
    $Result_i$ & Result Block & An intermediate logical block composed of one or more `B'-forms and `S'-forms. \\
    $RT$ & Relation Table & A truth value sequence derived from the SBE's operator structure. \\
    \bottomrule
  \end{tabularx}
\end{table}

\subsubsection{Decomposition into Structural Forms}
\leavevmode\par
First, the algorithm recursively decomposes the input SBE into a combination of the \textbf{\textit{B-forms}} and \textbf{\textit{S-forms}} defined in \textbf{Table~\ref{tab:notation}}. For example, an SBE such as `(X1 \&\& X2) || X3' is recognized as a structure where one `B'-form ($B_1^\wedge = X1 \wedge X2$) and one `S'-form ($S_2 = X3$) are combined by an OR operator, represented as $B_1^\wedge \vee S_2$. This step transforms a complex conditional expression into a standardized combination of basic units that the algorithm can process systematically.

\subsubsection{Relation Table Generation}
\leavevmode\par
Once the structural decomposition is complete, the algorithm analyzes the hierarchy of operators in the decomposed structure to generate a \textbf{\textit{Relation Table (RT)}}. The purpose of this table is to convert the complex, nested operator relationships into a simple, linear sequence of truth values that can be used in Phase 4. For instance, for a structure $B_1^\wedge \wedge B_2^\vee$, the algorithm identifies the operator combination ($\wedge, \wedge, \vee$) and, according to predefined rules, converts it into a final truth value sequence, `RT(T, T, F)'. This table acts as a clear \textbf{blueprint} that dictates which base patterns to apply and in what order during the test case composition.


\subsection{Phase 4: Test Case Generation}
In the final phase, the empty table created in Phase 2 is populated with truth values according to the analysis rules established in Phase 3. The core generation mechanism of Robin's Rule works by analyzing the structure of the SBE and applying the corresponding generation rule. This overall workflow is described in the generalized form in \textbf{Algorithm~\ref{alg:al3}}, where each branch (if-elseif-else) is detailed in the subsequent subsections 3.4.1, 3.4.2, and 3.4.3.


\begin{algorithm}[h!]
\caption{Generalized Test Case Generation Procedure based on the Relation Table (RT)}\label{alg:al3}
\begin{algorithmic}[1]
\State \textbf{Input:} $S_{norm}$ (A preprocessed SBE), $RT$ (The Relation Table from Phase 3).
\State \textbf{Output:} $T_{final}$ (The final $(N+1) \times N$ test case table).

\State $SubExprs \gets \textsc{DecomposeSBE}(S_{norm})$
\State $N \gets \textsc{CountOperands}(S_{norm})$
\State $T_{final} \gets \textsc{InitializeTable}(N+1, N)$
\State $T_{current} \gets \textsc{GetBasePatterns}(SubExprs.pop())$ \Comment{Initialize with the first block}

\Statex \Comment{Iterate through SBE blocks, referencing pre-generated RT to expand the table}
\While{$SubExprs$ is not empty}
    \State $NextBlock \gets SubExprs.pop()$
    \State $Op \gets \textsc{GetOperatorBetween}(T_{current}, NextBlock)$

    \Statex \Comment{Table combination and expansion using RT}
    \If{$NextBlock$ is a B-Form}
        \State $P_{next} \gets \textsc{GetBasePatterns}(NextBlock)$
        \State $T_{expanded} \gets \textsc{LayoutTables}(T_{current}, P_{next})$
        \State $\textsc{FillEmptyRegions}(T_{expanded}, RT)$ \Comment{Use RT per Sec. 3.4.1}
    \ElsIf{$NextBlock$ is an S-Form}
        \State $T_{expanded} \gets \textsc{LayoutTables}(T_{current}, \text{S\_Column})$
        \State $\textsc{FillSColumn}(T_{expanded}, RT, Op)$ \Comment{Use RT per Sec. 3.4.2}
    \EndIf

    \Statex \Comment{Handle exceptional cases during composition using RT}
    \If{\textsc{IsExceptionCase}($T_{expanded}$)}
        \State $T_{corrected} \gets \textsc{HandleException}(T_{expanded}, RT)$\Comment{Use RT per Sec. 3.4.3}
        \State $T_{current} \gets T_{corrected}$
    \Else
        \State $T_{current} \gets T_{expanded}$
    \EndIf
\EndWhile

\State $T_{final} \gets T_{current}$
\State \textbf{return} $T_{final}$
\end{algorithmic}
\end{algorithm}


\subsubsection{Test Case Generation for the `B o B' Form}
\leavevmode\par
For an SBE with a 4-condition (N=4) structure where two `B'-forms are combined, such as $B_i^o \circ B_{i+1}^o$, the algorithm generates a total of $N+1=5$ test cases. The entire process is illustrated in \textbf{Figure~\ref{fig:process}} and consists of the following sub-steps:

\begin{figure*}[t]
    \centering
    \includegraphics[width=\textwidth]{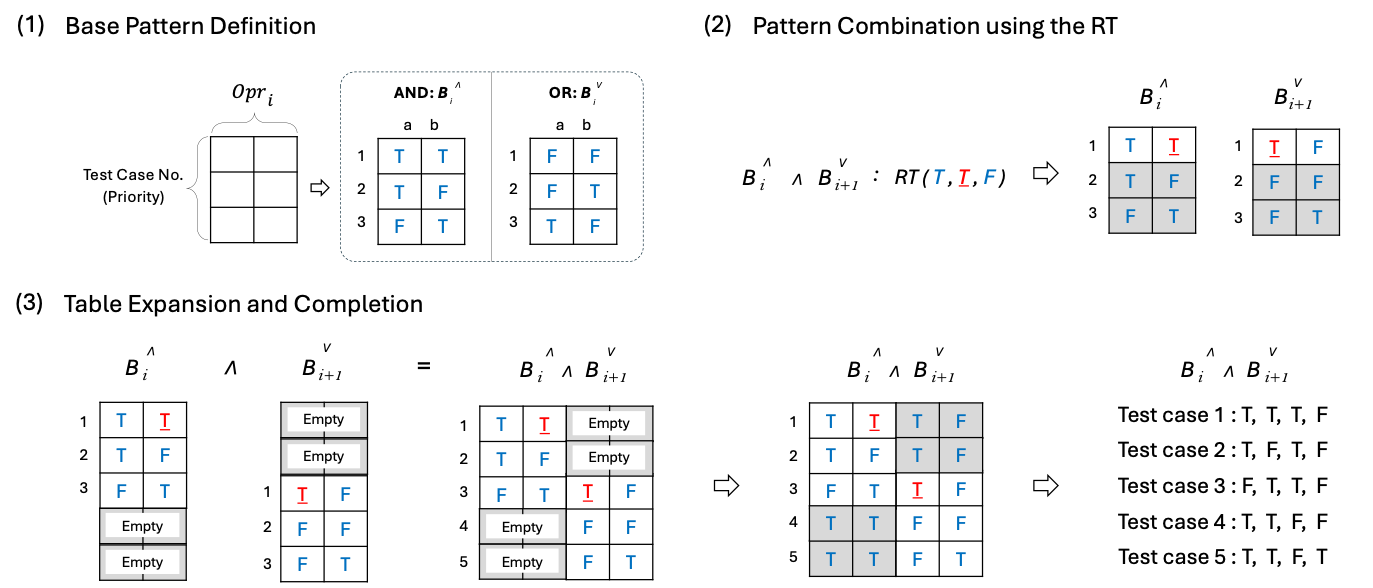}
    \caption{The overall three-step process for generating N+1 test cases from a `B o B' form.}
    \label{fig:process}
\end{figure*}

\begin{enumerate}
    \item \textbf{Base Pattern Definition}: First, the algorithm utilizes predefined 3-row `Base Pattern Tables' for the fundamental 2-condition `B'-form. These patterns are fixed based on operator priority to satisfy Unique-Cause MC/DC. For the `AND' operator, the minimal patterns are (T, T), (T, F), and (F, T). For the `OR' operator, the patterns are (F, F), (F, T), and (T, F). These serve as the essential building blocks for all test case combinations.
    
    \item \textbf{Pattern Combination using the RT}: The algorithm references the RT to combine the base patterns of the two `B'-forms. For example, let's assume the RT derived from the structure $B_i^\wedge \wedge B_{i+1}^\vee$ is (T, T, F). The algorithm maps the first two values (T, T) to $B_i^\wedge$ and the last two values (T, F) to $B_{i+1}^\vee$ to select their respective initial patterns. In this case, (T, T) is selected for $B_i^\wedge$, and (T, F) is selected for $B_{i+1}^\vee$. Afterward, the remaining areas are filled according to the priority test cases defined in (1) Base Pattern Definition.
    
    \item \textbf{Table Expansion and Completion}: The two selected base patterns are used to construct the larger 5x4 test table. To create the final 5-row table, the pattern table for $B_{i+1}^\vee$ is placed two rows down relative to the table for $B_i^\wedge$. The resulting empty regions are then filled deterministically based on the values from the RT(T, T, F). Specifically, the two empty cells in rows 4 and 5 corresponding to $B_i^\wedge$ are populated with (T, T), the first two values of the RT. Likewise, the two empty cells in rows 4 and 5 for $B_{i+1}^\vee$ are populated with (T, F), the last two values of the RT. This systematic process ensures that a complete and minimal set of 5 ($N+1$) test cases, which verifies the independence of all four conditions, is generated without duplication or omission.
\end{enumerate}


\subsubsection{Test Case Generation for the `B o S' Form}
\leavevmode\par
\begin{figure}[h!]
  \centering
  \includegraphics[width=\columnwidth]{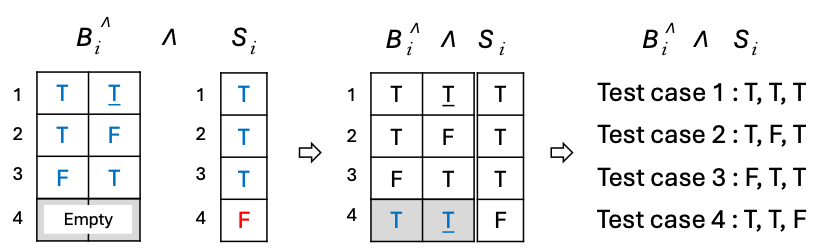} %
  \caption{Test case generation process for the `B o S' form.}
  \label{fig:b_o_s_generation}
\end{figure}

The process for a 3-condition (N=3) SBE combining a `B'-form and a single `S'-form differs based on the connecting operator. The core logic is summarized in \textbf{Figure~\ref{fig:b_o_s_generation}}.

\begin{enumerate}
    \item \textbf{Case of AND combination (e.g., $B_i^\wedge \wedge S_i$):} As illustrated in the examples in \textbf{Figure~\ref{fig:b_o_s_generation}}, this structure may follow a RT of (T, T). The algorithm combines the base pattern table (3x2) for $B_i^\wedge$ and the pattern (1-column) for $S_i$. The core rule is as follows: for the first $N$ test cases, the $S_i$ column is populated entirely with T, and only in the final, $N+1$th test case is the value inverted to F to prove the independence of $S_i$.

    \item \textbf{Case of OR combination (e.g., $B_i^\wedge \vee S_i$):}
    When combined with OR, the RT changes to (T, F), and the method for filling the $S_i$ column is the opposite of the AND case. That is, for the first $N$ test cases, the $S_i$ column is populated entirely with F, and only in the final, $N+1$th test case is the value inverted to T to prove the independence of $S_i$.
\end{enumerate}

\subsubsection{Exception Handling}
\leavevmode\par
For complex SBEs with deeply nested logical blocks, the simple composition rules described previously may be insufficient to guarantee the generation of a minimal $N+1$ test set. Specifically, Exceptional Cases can occur where invalid or duplicate test cases are generated.

\begin{figure}[h!]
   \centering
   \includegraphics[width=\columnwidth]{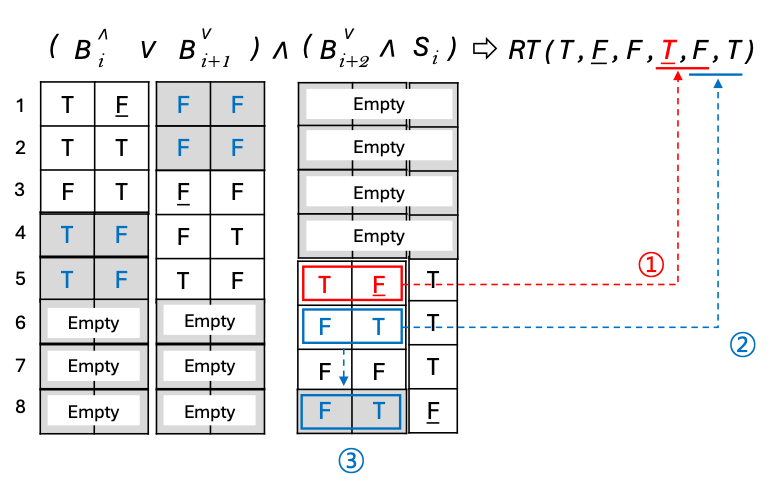}
   \caption{Application of the exception handling rule.}
   \label{fig:exception_handling_example}
\end{figure}

\begin{figure}[h!]
   \centering
   \includegraphics[width=\columnwidth]{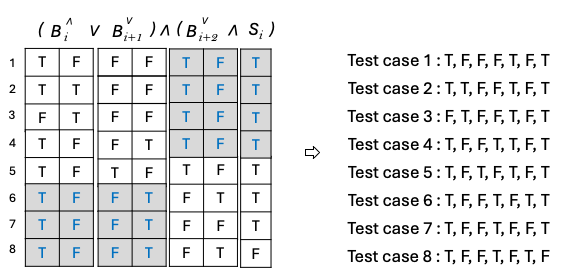}
   \caption{Test case generation}
   \label{fig:exception_case}
\end{figure}

\begin{enumerate}
    \item\textbf{{Defining the Exception Scenario}}\par
    An exception can arise when two or more previously generated intermediate Result Blocks, namely $Result_1$ and $Result_2$, are themselves combined by a logical operator. For example, an exception scenario occurs in a complex structure like the following:
    \begin{flalign*}\label{eq:complex_sbe_example}
        (B_i^o \circ B_{i+1}^o) \circ (B_{i+2}^o \circ S_{i}) \dots
    \end{flalign*} 
    In this structure, the algorithm broadly recognizes the expression as two intermediate result blocks, such as $Result_1 = (B_i^o \circ B_{i+1}^o)$ and $Result_2 = (B_{i+2}^o \circ S_{i})\dots$ At this final combination point (the `Handling Exception' point), the test patterns generated independently for each `Result' block can conflict, leading to a violation of the $N+1$ rule or causing redundancy.

    \item\textbf{Method for Handling Exceptions: Analysis of the \textbf{Figure~\ref{fig:exception_handling_example}} Example}\par
    To handle such exceptions, Robin's Rule applies a predefined rule illustrated in \textbf{Figure~\ref{fig:exception_handling_example}}.
    This figure illustrates the test table generation process for a complex SBE like $({B_{i}}^{\wedge}\vee{B^{\vee}}_{i+1})\wedge({B^{\vee}}_{i+2}\wedge S_{i})$.
    \begin{itemize}
        \item Here, the RT is derived as (T, F, F, T, F, T). As can be seen at point \ding{172} in the figure, a test case duplication occurs when attempting to populate the 5th row using simple composition (i.e., filling with F, T for the `B o S' form).
        \item At this moment, the exception handling rule is applied. The arrow at \ding{172} first fills the priority pattern (T, F) from the last three values of the RT (T, F, T). Then, at the point indicated by arrow \ding{173}, row 6 is filled with the `B o S' form's RT (F, T), and row 7 is filled with (F, F).
        \item Point \ding{174} shows that the `B o S' form's RT (F, T) is filled. Finally, as shown in \textbf{Figure~\ref{fig:exception_case}}, the empty region in the bottom-left (rows 6, 7, 8) is populated with the four values from the left side of the RT (T, F, F, T), and the empty region in the top-right (rows 1, 2, 3, 4) is populated with the three values from the right side of the RT (T, F, T) to complete the final test set.
    \end{itemize}
    
\end{enumerate}

\subsection{Complexity Analysis}

Let $N$ be the number of operands (conditions) in an input SBE. Robin’s Rule deterministically constructs an $(N+1)$×$N$ test table, and its overall runtime is dominated by the cost of filling this table.

\textbf{Phase 1 (Preprocessing)} consists of applying De Morgan’s laws recursively to push NOT operators down to operands and then performing structural sorting. The De Morgan transformation visits each operator/operand in the expression tree at most a constant number of times; therefore, it is linear in the expression size. Structural sorting establishes a deterministic processing order among blocks; even if sorting is implemented using a comparison-based method, it is at most O($N$ log $N$) with respect to the number of operands/blocks and is asymptotically dominated by later table-construction costs.

\textbf{Phase 2 (Temporary Table Generation)} initializes a 2D table with $(N+1)$ rows and $N$ columns, which requires O($N^2$) space allocation.

\textbf{Phase 3 (Structural Analysis of SBE)} decomposes the normalized SBE into B-forms and S-forms and generates the Relation Table (RT). Decomposition traverses the expression structure once to identify forms, and RT generation scans the resulting operator hierarchy to produce a truth-value sequence. Both operations are linear in the number of operands/blocks, i.e., O($N$).

\textbf{Phase 4 (Test Case Generation)} performs iterative table composition according to \textbf{Algorithm 3}. At each composition step, the algorithm expands and fills a portion of the $(N+1)$×$N$ table by (i) laying out the current partial table with the next block’s pattern and (ii) filling the empty regions using RT. Across the entire execution, each cell of the final $(N+1)$×$N$ table is assigned a truth value a constant number of times (once when it is first filled, and possibly once more in exceptional-case handling). Therefore, the cumulative cost of FillEmptyRegions/FillSColumn/HandleException is proportional to the number of cells, i.e., O($N^2$).

In summary, the overall time complexity is O($N^2$), and the space complexity is O($N^2$), mainly due to the explicit construction of the $(N+1)$×$N$ test table. This polynomial complexity stands in contrast to truth-table-based approaches requiring $2^N$ combinations.


\section{Experimental Evaluation}\label{sec:evaluation}

To evaluate the correctness, minimality, and efficiency of the proposed algorithm, \textbf{`Robin's Rule'}, we conducted an experiment designed to answer three key Research Questions (RQs). This section details our experimental setup, research questions, and an analysis of the results.


\subsection{Experimental Setup}\par


\subsubsection{Benchmark Dataset}
\leavevmode\par
This experiment is based on the Traffic Collision Avoidance System II (TCAS-II) specification, a well-known and complex real-world benchmark for safety-critical systems \cite{Weyuker1994}. As the proposed algorithm is optimized for SBEs, we constructed our benchmark dataset by modifying the 15 core logical expressions from the original TCAS-II benchmark into SBE form, and adding 10 randomly generated SBE conditions (with random N and random usage of \&\&, ||, !, and parentheses) to broaden diversity. This transformation is based on the fact that the vast majority (99.7\%) of conditional statements in large-scale avionics software are SBEs \cite{Chilenski2001}. A comprehensive benchmark suite with a number of conditions ($N$) ranging from 5 to 26 was configured, as detailed in \textbf{Table~\ref{tab:benchmark}}. SBEs No. 1-15 are reformulated from TCAS-II, while No. 16-25 are synthetically generated SBEs to introduce additional structural variety. The full textual forms of all 25 SBE expressions are provided in \textbf{Appendix~A} to improve readability while preserving reproducibility.


\begin{table}[t]
  \centering
  \caption{SBE benchmark specifications (full expressions are provided in Appendix~A).}
  \label{tab:benchmark}
  \scriptsize
  \setlength{\tabcolsep}{8pt}
  \renewcommand{\arraystretch}{0.9}
  \begin{tabular}{c c c c}
    \toprule
    \textbf{No} & \textbf{Origin} & \textbf{Condition (N)} & \textbf{Full SBE} \\
    \midrule
    1  & TCAS-II & 23 & App.~A, Listing~\ref{lst:sbe1}  \\
    2  & TCAS-II & 5  & App.~A, Listing~\ref{lst:sbe2}  \\
    3  & TCAS-II & 20 & App.~A, Listing~\ref{lst:sbe3}  \\
    4  & TCAS-II & 21 & App.~A, Listing~\ref{lst:sbe4}  \\
    5  & TCAS-II & 17 & App.~A, Listing~\ref{lst:sbe5}  \\
    6  & TCAS-II & 10 & App.~A, Listing~\ref{lst:sbe6}  \\
    7  & TCAS-II & 15 & App.~A, Listing~\ref{lst:sbe7}  \\
    8  & TCAS-II & 20 & App.~A, Listing~\ref{lst:sbe8}  \\
    9  & TCAS-II & 17 & App.~A, Listing~\ref{lst:sbe9}  \\
    10 & TCAS-II & 13 & App.~A, Listing~\ref{lst:sbe10} \\
    11 & TCAS-II & 12 & App.~A, Listing~\ref{lst:sbe11} \\
    12 & TCAS-II & 18 & App.~A, Listing~\ref{lst:sbe12} \\
    13 & TCAS-II & 11 & App.~A, Listing~\ref{lst:sbe13} \\
    14 & TCAS-II & 9  & App.~A, Listing~\ref{lst:sbe14} \\
    15 & TCAS-II & 8  & App.~A, Listing~\ref{lst:sbe15} \\
    \midrule
    16 & Random  & 19 & App.~A, Listing~\ref{lst:sbe16} \\
    17 & Random  & 11 & App.~A, Listing~\ref{lst:sbe17} \\
    18 & Random  & 24 & App.~A, Listing~\ref{lst:sbe18} \\
    19 & Random  & 18 & App.~A, Listing~\ref{lst:sbe19} \\
    20 & Random  & 26 & App.~A, Listing~\ref{lst:sbe20} \\
    21 & Random  & 13 & App.~A, Listing~\ref{lst:sbe21} \\
    22 & Random  & 17 & App.~A, Listing~\ref{lst:sbe22} \\
    23 & Random  & 7  & App.~A, Listing~\ref{lst:sbe23} \\
    24 & Random  & 11 & App.~A, Listing~\ref{lst:sbe24} \\
    25 & Random  & 18 & App.~A, Listing~\ref{lst:sbe25} \\
    \bottomrule
  \end{tabular}
\end{table}


\subsubsection{Experimental Procedure}
\leavevmode\par
The experiment was conducted according to the following systematic procedure:
\begin{enumerate}
    \item \textbf{Source Code Generation:} For each of the 25 SBEs in the benchmark dataset, a C language source file was automatically generated, containing the respective SBE within an `if' statement.
    \item \textbf{Test Case Generation:} Using the implemented Robin's Rule tool, each C source file was analyzed to generate a set of test cases satisfying Unique-Cause MC/DC, which was then saved in CSV file format.
    \item \textbf{Coverage Verification:} The generated CSV files were imported into the VectorCAST test environment, where the tests were executed and the results were used to measure the Unique-Cause MC/DC coverage.
\end{enumerate}


\subsubsection{Validation Tool and Comparison Target}
\leavevmode\par
To validate the correctness and coverage of the generated test cases, we selected VectorCAST (ver. 2025) as an oracle. VectorCAST is an industry-standard commercial tool qualified for avionics standards such as DO-178C \cite{Santhanam2007,DO330,Godboley2021a}. Therefore, its analysis results can serve as a reliable oracle for validating the achieved Unique-Cause MC/DC coverage.

In addition to this coverage-oracle validation, we compare our approach against two representative baseline paradigms used in state-of-the-art MC/DC test generation for automated Boolean reasoning in MC/DC test generation: (i) a BDD(roBDD)-based generator, following the roBDD-based MC/DC generation line \cite{Ahishakiye2021}, and (ii) a SAT-solver-based generator, following SAT-based MC/DC test generation approaches \cite{Kitamura2018}. We use Z3 as the backend solver for the SAT baseline due to its maturity and widespread adoption in satisfiability-based program analysis.

These baselines are complementary: BDD-based methods construct a canonical decision-graph representation that supports systematic extraction of distinguishing input assignments, while SAT-based methods encode coverage obligations as satisfiability constraints and leverage solver technology to compute satisfying assignments. Together, they provide strong and informative references for assessing whether our direct-construction approach offers advantages in minimality, efficiency, and coverage guarantees.

We report (i) the number of generated test cases, (ii) generation time, and (iii) achieved Unique-Cause MC/DC coverage across all 25 SBEs (\textbf{Tables~\ref{tab:condition}–\ref{tab:results}}).


\subsection{Research Questions (RQs)}\par
This evaluation was structured to answer the following research questions:
\begin{itemize}
    \item[\textbf{RQ1}] \textbf{(Minimality):} Can Robin's Rule consistently generate a minimal test set of size $N+1$ for SBEs with $N$ conditions?
    \item[\textbf{RQ2}] \textbf{(Correctness):} Do the $N+1$ test cases generated by our algorithm achieve 100\% Unique-Cause MC/DC coverage when verified using VectorCAST as an independent coverage oracle?
    \item[\textbf{RQ3}] \textbf{(Comparison with Baselines):} How does Robin’s Rule compare to BDD and SAT-based baselines in terms of (a)test suite size, (b) generation time, and (c) achieved Unique-Cause MC/DC coverage?
\end{itemize}


\subsection{Case Study: Detailed Analysis of SBE No. 1}\par

This section describes how Robin's Rule generates a minimal test set, using a representative large-scale benchmark, SBE No. 1 (N=23), as a target.


\subsubsection{Phase 1 \& 2 Application for SBE No. 1}
\leavevmode\par
When the original conditional expression of SBE No. 1 undergoes the Preprocessing phase, it is transformed into the following normalized and sorted structure:

\begin{lstlisting}
(((((n||o)&&m)&&l)&&p)||(((r||s)&&q)&&!t)||((v||w)
&&u))&&(((c&&!d)&&!e)||((!f&&g)&&!h)||((!i&&!j)
&&!k))&&(!a||!b)
\end{lstlisting}

As there are 23 conditions, a temporary table of size 24 rows by 23 columns is generated.


\subsubsection{Test Case Generation Results}
\leavevmode\par
In Phase 3, the Relation Table (RT) is derived as \textbf{RT(F, T, T, T, F, F, T, T, F, F, T, T, T, T, F, T, T, F, T, T, T, F)}. Then, in Phase 4, Robin's Rule uses this RT to generate the final set of 24 ($N+1$) test cases for SBE No. 1, which is the theoretical minimum, as shown in \textbf{Table~\ref{tab:sbe_testcases}}. Each row represents a single test case, and each column represents the truth value (1=True, 0=False) for each condition from `a' to `w'.


\begin{table}[h!]
\centering
\caption{The 24 Test Cases Generated by Robin's Rule for SBE No. 1 ($N$=23).}
\label{tab:sbe_testcases}
\small 
\setlength{\tabcolsep}{1.6pt} 
\renewcommand{\arraystretch}{0.85}

\begin{tabular}{c|cccccccccccccccccccccccc}
\toprule
 \textbf{Test Case} & \textbf{n} & \textbf{o} & \textbf{m} & \textbf{l} & \textbf{p} & \textbf{r} & \textbf{s} & \textbf{q} & \textbf{!t} & \textbf{v} & \textbf{w} & \textbf{u} & \textbf{c} & \textbf{!d} & \textbf{!e} & \textbf{!f} & \textbf{g} & \textbf{!h} & \textbf{!i} & \textbf{!j} & \textbf{!k} & \textbf{!a} & \textbf{!b} \\
\midrule
\textbf{ 1} & 0 & 1 & 1 & 1 & 1 & 0 & 0 & 1 & 0 & 0 & 0 & 1 & 1 & 0 & 0 & 1 & 1 & 0 & 1 & 0 & 0 & 0 & 1 \\
\textbf{ 2} & 0 & 0 & 1 & 1 & 1 & 0 & 0 & 1 & 0 & 0 & 0 & 1 & 1 & 0 & 0 & 1 & 1 & 0 & 1 & 0 & 0 & 0 & 1 \\
\textbf{ 3} & 1 & 0 & 1 & 1 & 1 & 0 & 0 & 1 & 0 & 0 & 0 & 1 & 1 & 0 & 0 & 1 & 1 & 0 & 1 & 0 & 0 & 0 & 1 \\
\textbf{ 4} & 0 & 1 & 0 & 1 & 1 & 0 & 0 & 1 & 0 & 0 & 0 & 1 & 1 & 0 & 0 & 1 & 1 & 0 & 1 & 0 & 0 & 0 & 1 \\
\textbf{ 5} & 0 & 1 & 1 & 0 & 1 & 0 & 0 & 1 & 0 & 0 & 0 & 1 & 1 & 0 & 0 & 1 & 1 & 0 & 1 & 0 & 0 & 0 & 1 \\
\textbf{ 6} & 0 & 1 & 1 & 1 & 0 & 0 & 0 & 1 & 0 & 0 & 0 & 1 & 1 & 0 & 0 & 1 & 1 & 0 & 1 & 0 & 0 & 0 & 1 \\
\textbf{ 7} & 0 & 1 & 1 & 1 & 0 & 0 & 1 & 1 & 0 & 0 & 0 & 1 & 1 & 0 & 0 & 1 & 1 & 0 & 1 & 0 & 0 & 0 & 1 \\
\textbf{ 8} & 0 & 1 & 1 & 1 & 0 & 1 & 0 & 1 & 0 & 0 & 0 & 1 & 1 & 0 & 0 & 1 & 1 & 0 & 1 & 0 & 0 & 0 & 1 \\
\textbf{ 9} & 0 & 1 & 1 & 1 & 0 & 0 & 1 & 0 & 0 & 0 & 0 & 1 & 1 & 0 & 0 & 1 & 1 & 0 & 1 & 0 & 0 & 0 & 1 \\
\textbf{10} & 0 & 1 & 1 & 1 & 0 & 0 & 1 & 1 & 1 & 0 & 0 & 1 & 1 & 0 & 0 & 1 & 1 & 0 & 1 & 0 & 0 & 0 & 1 \\
\textbf{11} & 0 & 1 & 1 & 1 & 0 & 0 & 1 & 1 & 1 & 0 & 1 & 1 & 1 & 0 & 0 & 1 & 1 & 0 & 1 & 0 & 0 & 0 & 1 \\
\textbf{12} & 0 & 1 & 1 & 1 & 0 & 0 & 1 & 1 & 1 & 1 & 0 & 1 & 1 & 0 & 0 & 1 & 1 & 0 & 1 & 0 & 0 & 0 & 1 \\
\textbf{13} & 0 & 1 & 1 & 1 & 0 & 0 & 1 & 1 & 1 & 0 & 1 & 0 & 1 & 0 & 0 & 1 & 1 & 0 & 1 & 0 & 0 & 0 & 1 \\
\textbf{14} & 0 & 1 & 1 & 1 & 0 & 0 & 1 & 1 & 1 & 0 & 1 & 1 & 1 & 1 & 0 & 1 & 1 & 0 & 1 & 0 & 0 & 0 & 1 \\
\textbf{15} & 0 & 1 & 1 & 1 & 0 & 0 & 1 & 1 & 1 & 0 & 1 & 1 & 0 & 0 & 0 & 1 & 1 & 0 & 1 & 0 & 0 & 0 & 1 \\
\textbf{16} & 0 & 1 & 1 & 1 & 0 & 0 & 1 & 1 & 1 & 0 & 1 & 1 & 1 & 0 & 1 & 1 & 1 & 0 & 1 & 0 & 0 & 0 & 1 \\
\textbf{17} & 0 & 1 & 1 & 1 & 0 & 0 & 1 & 1 & 1 & 0 & 1 & 1 & 1 & 0 & 1 & 0 & 1 & 0 & 1 & 0 & 0 & 0 & 1 \\
\textbf{18} & 0 & 1 & 1 & 1 & 0 & 0 & 1 & 1 & 1 & 0 & 1 & 1 & 1 & 0 & 1 & 0 & 0 & 0 & 1 & 0 & 0 & 0 & 1 \\
\textbf{19} & 0 & 1 & 1 & 1 & 0 & 0 & 1 & 1 & 1 & 0 & 1 & 1 & 1 & 0 & 1 & 0 & 1 & 1 & 1 & 0 & 0 & 0 & 1 \\
\textbf{20} & 0 & 1 & 1 & 1 & 0 & 0 & 1 & 1 & 1 & 0 & 1 & 1 & 1 & 0 & 1 & 0 & 1 & 1 & 0 & 0 & 0 & 0 & 1 \\
\textbf{21} & 0 & 1 & 1 & 1 & 0 & 0 & 1 & 1 & 1 & 0 & 1 & 1 & 1 & 0 & 1 & 0 & 1 & 1 & 0 & 1 & 0 & 0 & 1 \\
\textbf{22} & 0 & 1 & 1 & 1 & 0 & 0 & 1 & 1 & 1 & 0 & 1 & 1 & 1 & 0 & 1 & 0 & 1 & 1 & 0 & 0 & 1 & 0 & 1 \\
\textbf{23} & 0 & 1 & 1 & 1 & 0 & 0 & 1 & 1 & 1 & 0 & 1 & 1 & 1 & 0 & 1 & 0 & 1 & 1 & 0 & 0 & 0 & 1 & 1 \\
\textbf{24} & 0 & 1 & 1 & 1 & 0 & 0 & 1 & 1 & 1 & 0 & 1 & 1 & 1 & 0 & 1 & 0 & 1 & 1 & 0 & 0 & 0 & 1 & 0 \\
\bottomrule
\end{tabular}
\end{table}


\subsection{Threats to Validity}\par

To ensure the academic rigor of our study, we identify and discuss the potential threats to the validity of our findings.

\paragraph{\textbf{(Construct Validity)}}\par
This threat relates to whether we are truly measuring what we claim to be measuring. Our study relies on VectorCAST, an industry-standard commercial tool, as the oracle for MC/DC coverage. While this tool is certified for DO-178C, which strengthens our confidence, there remains a theoretical possibility that its specific implementation or interpretation of the Unique-Cause criterion could influence the results. To mitigate this, we chose a tool that is widely accepted and trusted in the safety-critical domain.

\paragraph{\textbf{(Internal Validity)}}\par
This threat concerns potential flaws within our own experimental setup. There could be an error in our implementation of the Robin's Rule algorithm. We mitigated this threat by cross-validating our results: for every SBE, the test set generated by our tool was independently verified using VectorCAST as an external commercial coverage-analysis tool to achieve 100\% coverage, which provides strong evidence for the correctness of our implementation.

\paragraph{\textbf{(External Validity)}}\par
This threat relates to the generalizability of our results. First, our benchmark, while based on the TCAS-II specification, consists of SBEs that were manually modified from the original expressions. It is possible that this transformation simplified some of the original logic. Second, our study is intentionally focused on SBEs. While our introduction provides a strong rationale for this focus (99.7\% prevalence), the results cannot be directly generalized to Non-SBEs containing coupled conditions. Extending the algorithm to handle Non-SBEs remains a crucial area for future work.


\subsection{Overall Results}\par

This section analyzes the experimental results for each research question (RQ) in sequence.

\paragraph{Answering RQ1: Generation of Minimal $N+1$ Test Sets}\par
To answer RQ1, we analyzed the number of test cases generated by Robin's Rule. For every SBE specification with $N$ conditions, our algorithm generated \textbf{exactly $N+1$ test cases}. This result was consistent across the entire benchmark, from $N$=5 to $N$=26, empirically confirming that Robin's Rule successfully generates a test set of the theoretical minimum size for Unique-Cause MC/DC.

\paragraph{Answering RQ2: Achievement of 100\% Coverage}\par
To answer RQ2, each of the $N+1$ test sets was validated using VectorCAST. The results show that \textbf{100\% Unique-Cause MC/DC coverage was achieved in all 25 cases}. This demonstrates the correctness and effectiveness of the proposed deterministic generation logic.

\paragraph{Answering RQ3: Comparison with BDD, SAT solver}\par
\begin{figure}[h!]
   \centering
   \includegraphics[width=\columnwidth]{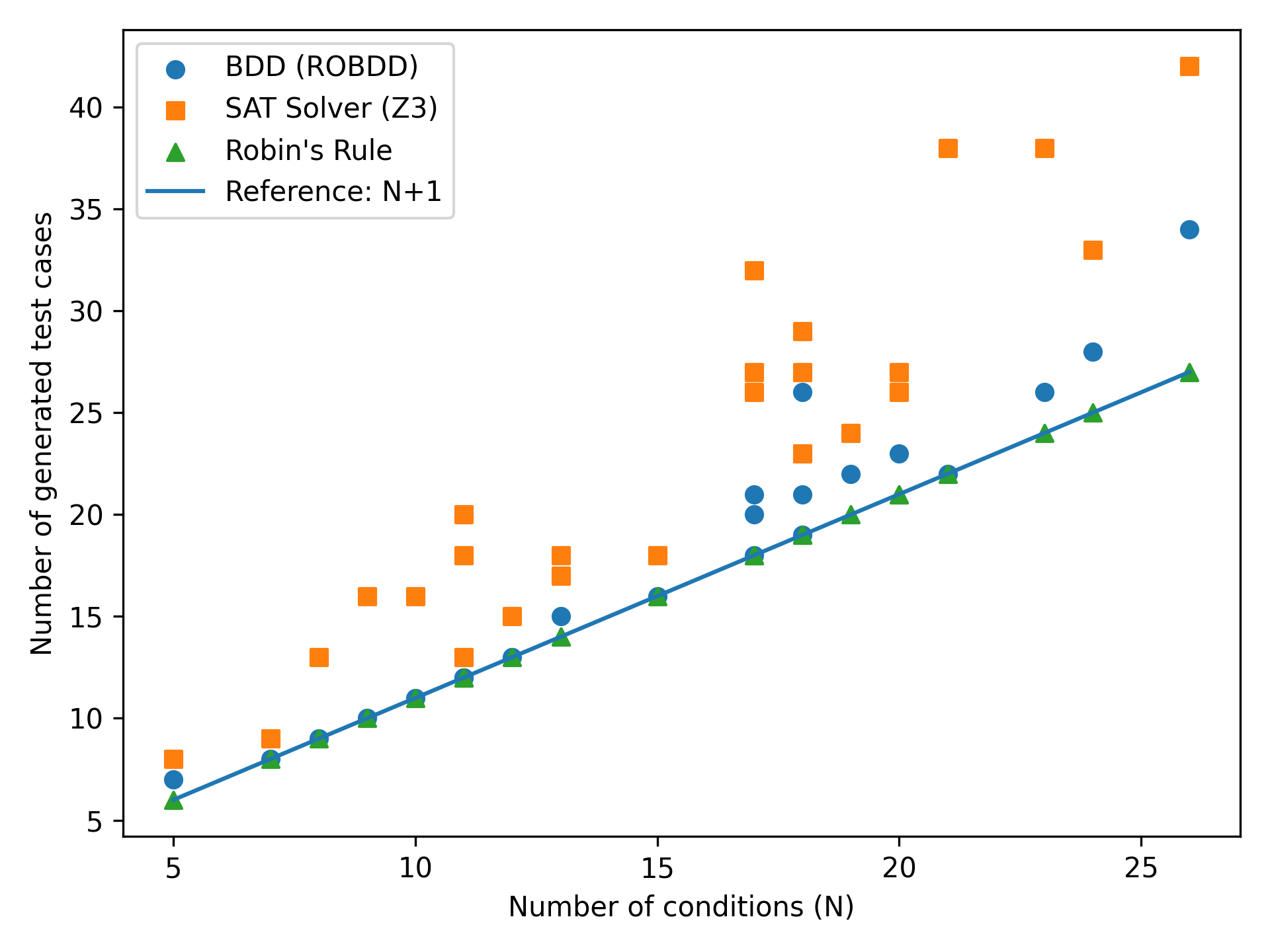}
   \caption{Number of generated test cases (BDD vs SAT vs Robin’s Rule)}
   \label{fig:condition}
\end{figure}

\begin{figure}[h!]
   \centering
   \includegraphics[width=\columnwidth]{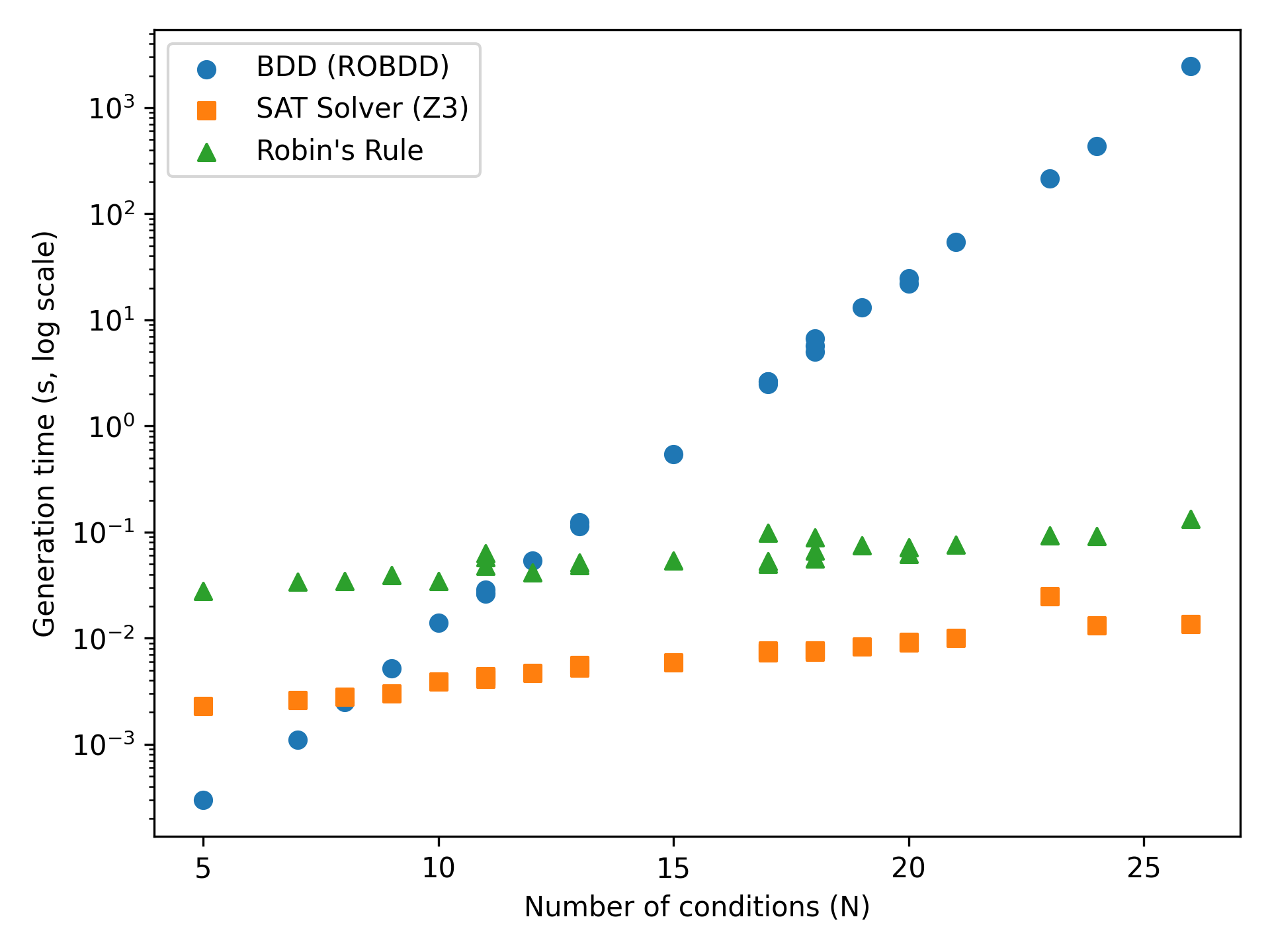}
   \caption{Test case generation time}
   \label{fig:time}
\end{figure}

\begin{figure}[h!]
   \centering
   \includegraphics[width=\columnwidth]{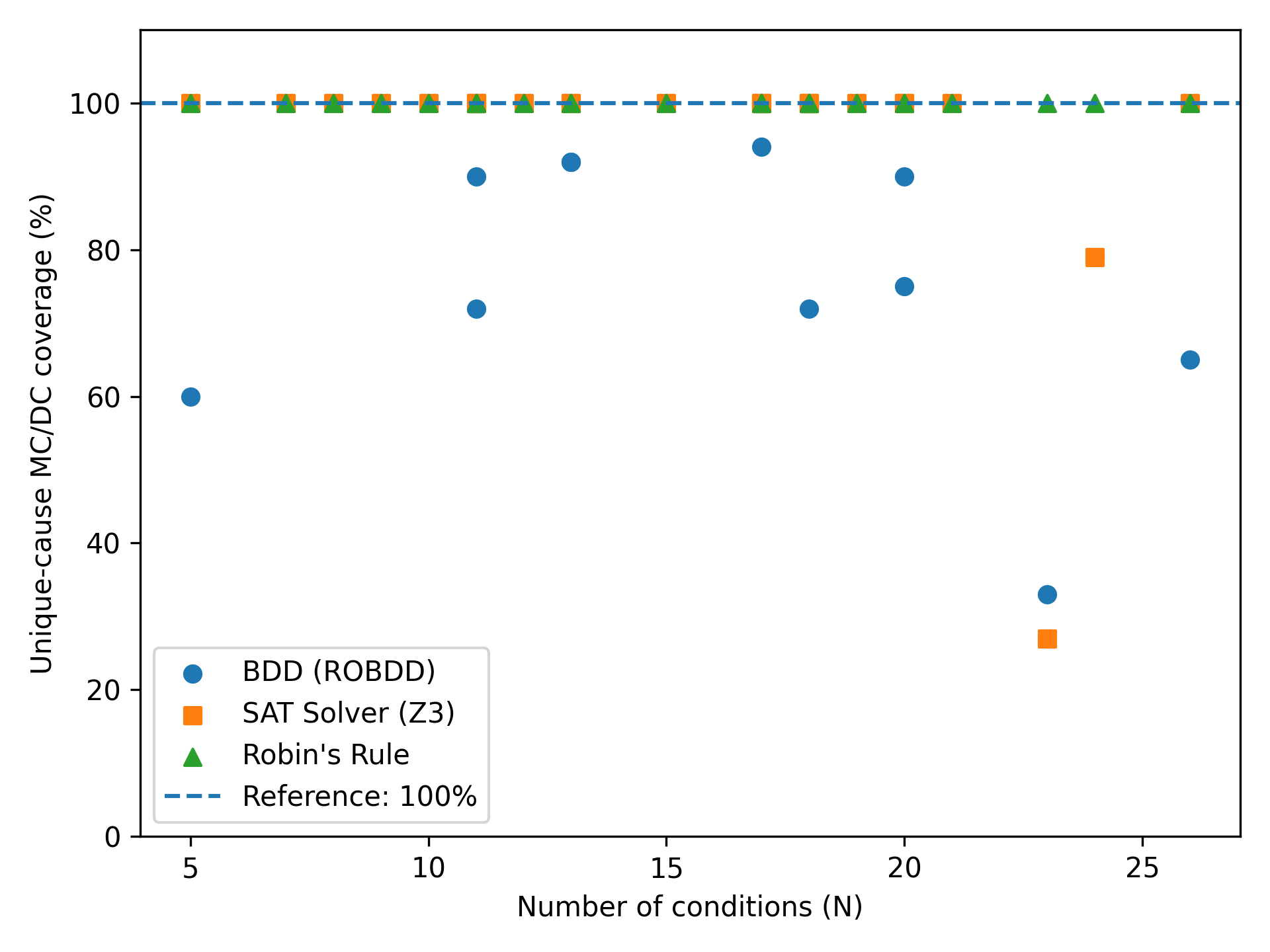}
   \caption{MC/DC test results}
   \label{fig:coverage}
\end{figure}

 To address RQ3, we compare Robin’s Rule against BDD(roBDD) and SAT(Z3) baselines. \textbf{Table~\ref{tab:condition}-\ref{tab:results}} summarize the numerical results, and Figures \ref{fig:condition}-\ref{fig:coverage} visualize the same outcomes.

\begin{itemize}
\item \textbf{Test suite size (Table~\ref{tab:condition} and Figure~\ref{fig:condition}):} Robin’s Rule consistently generates exactly $N+1$ test cases for every SBE, while BDD and SAT often require more than $N+1$.
\item \textbf{Generation time (Table~\ref{tab:time} and Figure~\ref{fig:time}):} SAT and Robin’s Rule remains stable as $N$ grows, whereas BDD shows a clear increasing trend in generation time with larger $N$.
\item \textbf{Coverage (Table~\ref{tab:results} and Figure~\ref{fig:coverage}):} Robin’s Rule achieves 100\% Unique-Cause MC/DC coverage for all SBEs, while the baselines do not always reach full coverage.
\end{itemize}

\textbf{Table~\ref{tab:condition} (Test suite size).}
Table~\ref{tab:condition} reports, for each of the 25 SBEs, the number of conditions $N$ and the resulting test-suite size produced by each generator. Robin’s Rule generates exactly $N{+}1$ test cases for all SBEs (ranging from 6 to 27 cases), matching the theoretical minimum. In contrast, the BDD baseline matches $N{+}1$ only in 11 out of 25 SBEs and otherwise requires additional tests (up to $+7$ over $N{+}1$, with an average overhead of $+1.68$). The SAT baseline exceeds $N{+}1$ in all 25 cases, requiring on average $+6.6$ additional tests and reaching the maximum overhead of $+16$ (SBE No.~4).

\begin{table}[h!]
  \centering
  \caption{Test Case Generation Results (Number of Test Cases)}
  \label{tab:condition}
  \small
  \setlength{\tabcolsep}{3pt}
  \renewcommand{\arraystretch}{0.85}
  \begin{tabular}{c|c|c|c|c}
    \toprule
    \textbf{SBE} & \textbf{No. of} & \multicolumn{1}{c|}{\textbf{BDD}} & \multicolumn{1}{c|}{\textbf{SAT}} & \multicolumn{1}{c}{\textbf{Robin's Rule}} \\
    \textbf{No.} & \textbf{Cond. ($N$)} & \textbf{Cases} & \textbf{Cases} & \textbf{Cases} \\
    \midrule
     1 & 23 & 26 & 38 & \textbf{24} \\
     2 &  5 &  7 &  8 & \textbf{6}  \\
     3 & 20 & 27 & 27 & \textbf{21} \\
     4 & 21 & 22 & 38 & \textbf{22} \\
     5 & 17 & 18 & 32 & \textbf{18} \\
     6 & 10 & 11 & 16 & \textbf{11} \\
     7 & 15 & 16 & 18 & \textbf{16} \\
     8 & 20 & 23 & 26 & \textbf{21} \\
     9 & 17 & 20 & 27 & \textbf{18} \\
    10 & 13 & 15 & 17 & \textbf{14} \\
    11 & 12 & 13 & 15 & \textbf{13} \\
    12 & 18 & 19 & 23 & \textbf{19} \\
    13 & 11 & 12 & 18 & \textbf{12} \\
    14 &  9 & 10 & 16 & \textbf{10} \\
    15 &  8 &  9 & 13 & \textbf{9}  \\
    16 & 19 & 22 & 24 & \textbf{20} \\
    17 & 11 & 12 & 13 & \textbf{12} \\
    18 & 24 & 28 & 33 & \textbf{25} \\
    19 & 18 & 26 & 27 & \textbf{19} \\
    20 & 26 & 34 & 42 & \textbf{27} \\
    21 & 13 & 17 & 18 & \textbf{14} \\
    22 & 17 & 21 & 26 & \textbf{18} \\
    23 &  7 &  8 &  9 & \textbf{8}  \\
    24 & 11 & 13 & 20 & \textbf{12} \\
    25 & 18 & 21 & 29 & \textbf{19} \\
    \bottomrule
  \end{tabular}
\end{table}

\textbf{Figure~\ref{fig:condition} (Test suite size vs.\ $N$).}
Figure~\ref{fig:condition} visualizes the Table~\ref{tab:condition} results as a function of $N$ (from $N{=}5$ to $N{=}26$). The Robin’s Rule curve overlaps the $N{+}1$ lower bound across the entire range, indicating consistent minimality regardless of expression size. The BDD curve stays relatively close to the lower bound for smaller $N$ but diverges for larger and structurally complex SBEs, whereas the SAT curve remains consistently above the bound, reflecting that the SAT-based construction in our baseline typically produces additional assignments beyond the minimum.

\textbf{Table~\ref{tab:time} (Generation time).}
Table~\ref{tab:time} reports the wall-clock time (seconds) required to generate test cases for each SBE. The SAT baseline is the fastest overall (median 0.0073\,s; range 0.0023-0.0247\,s). Robin’s Rule remains stable and efficient (median 0.0565\,s; range 0.0280-0.1326\,s). In contrast, the BDD baseline exhibits a heavy-tailed distribution (median 2.468\,s; range 0.0003-2448.046\,s), including extreme outliers such as SBE No.~20 ($N{=}26$), which requires 2448.046\,s.

\begin{table}[h!]
  \centering
  \caption{Test Case Generation Time (s)}
  \label{tab:time}
  \small
  \setlength{\tabcolsep}{3pt}
  \renewcommand{\arraystretch}{0.85}
  \begin{tabular}{c|c|c|c}
    \toprule
    \textbf{SBE} & \multicolumn{1}{c|}{\textbf{BDD}} & \multicolumn{1}{c|}{\textbf{SAT}} & \multicolumn{1}{c}{\textbf{Robin's Rule}} \\
    \textbf{No.} & \textbf{Time (s)} & \textbf{Time (s)} & \textbf{Time (s)} \\
    \midrule
     1 & 215.9581 & 0.0247 & \textbf{0.092557} \\
     2 & 0.0003   & 0.0023 & \textbf{0.028010} \\
     3 & 22.0323  & 0.0091 & \textbf{0.062134} \\
     4 & 54.5546  & 0.01   & \textbf{0.076227} \\
     5 & 2.6119   & 0.0074 & \textbf{0.053130} \\
     6 & 0.01397  & 0.0039 & \textbf{0.034554} \\
     7 & 0.541    & 0.0059 & \textbf{0.054029} \\
     8 & 24.531   & 0.0092 & \textbf{0.071639} \\
     9 & 2.6163   & 0.0076 & \textbf{0.098991} \\
    10 & 0.1128   & 0.0056 & \textbf{0.049013} \\
    11 & 0.054    & 0.0047 & \textbf{0.041849} \\
    12 & 5.009    & 0.0077 & \textbf{0.056544} \\
    13 & 0.0265   & 0.0044 & \textbf{0.057589} \\
    14 & 0.0052   & 0.003  & \textbf{0.039289} \\
    15 & 0.0025   & 0.0028 & \textbf{0.034258} \\
    16 & 13.1345  & 0.0083 & \textbf{0.074857} \\
    17 & 0.0287   & 0.0043 & \textbf{0.048177} \\
    18 & 434.7779 & 0.0131 & \textbf{0.091852} \\
    19 & 6.6683   & 0.0076 & \textbf{0.089465} \\
    20 & 2448.046 & 0.0135 & \textbf{0.132568} \\
    21 & 0.1241   & 0.0053 & \textbf{0.051261} \\
    22 & 2.468    & 0.0073 & \textbf{0.050424} \\
    23 & 0.0011   & 0.0026 & \textbf{0.034107} \\
    24 & 0.0263   & 0.0041 & \textbf{0.062772} \\
    25 & 5.7416   & 0.0074 & \textbf{0.067168} \\
    \bottomrule
  \end{tabular}
\end{table}

\textbf{Figure~\ref{fig:time} (Time vs.\ $N$).}
Figure~\ref{fig:time} plots the same timing results, using a logarithmic scale to accommodate the large variance of BDD. As $N$ increases, SAT and Robin’s Rule stay within a narrow band (sub-second and stable), while BDD shows both a clear upward trend and high variance. This highlights the practical sensitivity of BDD-based generation to the expression structure and the resulting diagram size, even within the SBE class.

\textbf{Table~\ref{tab:results} (Achieved Unique-Cause MC/DC coverage).}
Table~\ref{tab:results} reports the achieved Unique-Cause MC/DC coverage validated by VectorCAST.
Robin's Rule achieves 100\% coverage for all 25 SBEs. The SAT baseline reaches 100\% in 23 out of 25 SBEs but drops to 27\% for SBE No.~1 and 79\% for SBE No.~18. The BDD baseline reaches 100\% in 13 out of 25 SBEs, with the minimum coverage being 33\% (SBE No.~1). These failures are consistent with masking effects caused by short-circuit semantics and the fact that solver/graph-based baselines do not uniformly enforce Unique-Cause obligations under a fixed implementation and configuration. Overall, the results indicate that, under our baseline implementations and settings, the solver/graph-based approaches do not provide a uniform guarantee of Unique-Cause MC/DC across all SBEs, whereas Robin's Rule does.

\begin{table}[h!]
  \centering
  \caption{Unique-Cause MC/DC Coverage Results (\%)}
  \label{tab:results}
  \small
  \setlength{\tabcolsep}{3pt}
  \renewcommand{\arraystretch}{0.85}
  \begin{tabular}{c|c|c|c}
    \toprule
    \textbf{SBE} & \multicolumn{1}{c|}{\textbf{BDD}} & \multicolumn{1}{c|}{\textbf{SAT}} & \multicolumn{1}{c}{\textbf{Robin's Rule}} \\
    \textbf{No.} & \textbf{Coverage (\%)} & \textbf{Coverage (\%)} & \textbf{Coverage (\%)} \\
    \midrule
     1 &  33\% &  27\% & \textbf{100\%} \\
     2 &  60\% & 100\% & \textbf{100\%} \\
     3 &  75\% & 100\% & \textbf{100\%} \\
     4 & 100\% & 100\% & \textbf{100\%} \\
     5 & 100\% & 100\% & \textbf{100\%} \\
     6 & 100\% & 100\% & \textbf{100\%} \\
     7 & 100\% & 100\% & \textbf{100\%} \\
     8 &  90\% & 100\% & \textbf{100\%} \\
     9 &  94\% & 100\% & \textbf{100\%} \\
    10 &  92\% & 100\% & \textbf{100\%} \\
    11 & 100\% & 100\% & \textbf{100\%} \\
    12 & 100\% & 100\% & \textbf{100\%} \\
    13 & 100\% & 100\% & \textbf{100\%} \\
    14 & 100\% & 100\% & \textbf{100\%} \\
    15 & 100\% & 100\% & \textbf{100\%} \\
    16 & 100\% & 100\% & \textbf{100\%} \\
    17 &  90\% & 100\% & \textbf{100\%} \\
    18 &  79\% &  79\% & \textbf{100\%} \\
    19 & 100\% & 100\% & \textbf{100\%} \\
    20 &  65\% & 100\% & \textbf{100\%} \\
    21 &  92\% & 100\% & \textbf{100\%} \\
    22 & 100\% & 100\% & \textbf{100\%} \\
    23 & 100\% & 100\% & \textbf{100\%} \\
    24 &  72\% & 100\% & \textbf{100\%} \\
    25 &  72\% & 100\% & \textbf{100\%} \\
    \bottomrule
  \end{tabular}
\end{table}

\textbf{Figure~\ref{fig:coverage} (Coverage visualization).}
Figure~\ref{fig:coverage} summarizes Table~\ref{tab:results} and highlights the key outcome: Robin’s Rule is the only approach that consistently attains full (100\%) Unique-Cause MC/DC coverage across all 25 SBEs, while the baselines exhibit non-trivial coverage gaps on specific SBEs.

Overall, these results indicate that Robin’s Rule simultaneously ensures minimality $(N+1)$, stable efficiency, and guaranteed 100\% Unique-Cause MC/DC coverage across the benchmark.


\subsection{Discussion}\par
The results have three main implications: optimality and correctness, efficiency in computational complexity, and deterministic reliability. Robin's Rule avoids the exponential cost of truth-table or exhaustive search methods by directly constructing the minimal set, and runs in O($N^2$) time by directly constructing and filling the $(N+1)$×$N$ test table. Its deterministic nature is a key advantage for tool qualification where predictability and reproducibility are paramount.
The primary threat to validity is the study's focus on SBEs. While SBEs are highly prevalent~\cite{Chilenski2001}, extending this work to Non-SBEs is necessary. Another limitation is the comparison against a single commercial tool. VectorCAST was chosen as our coverage oracle to validate correctness, and a broader comparative study against other tools remains an area for future work.


\section{Conclusion and Future Work}
This paper presented Robin's Rule, a deterministic algorithm that directly constructs a minimal $N+1$ test set for SBEs to satisfy Unique-Cause MC/DC. Experiments on a TCAS-II-based benchmark, validated using VectorCAST as an independent coverage oracle, demonstrated that our method consistently achieves 100\% coverage with the theoretical minimum number of test cases. This work offers a practical and optimal solution for developers in safety-critical domains, enabling rigorous software verification at a reduced cost. Future work will focus on extending the algorithm to handle Non-SBEs (coupled conditions) and adapting it for the Masking MC/DC environment.




\begin{thebibliography}{99}

\bibitem{Martins2020}
L. E. G. Martins and T. Gorschek. 2020. Requirements Engineering for Safety-Critical Systems: An Interview Study with Industry Practitioners. \textit{IEEE TSE}, 46(4): 346--361.

\bibitem{An2021}
H. C. An and H. S. Yang. 2021. Fully Adaptive Stochastic Handling of Soft-Errors in Real-Time Systems. \textit{IEEE Access}, 155058--155071.

\bibitem{Singh2021}
P. Singh and L. K. Singh. 2021. Reliability and Safety Engineering for Safety Critical Systems: An Interview Study With Industry Practitioners. \textit{IEEE Transactions on Reliability}, 70(2): 643--653.

\bibitem{Singh2018}
L. K. Singh and H. Rajput. 2018. Dependability Analysis of Safety Critical Real-Time Systems by Using Petri Nets. \textit{IEEE Transactions on Control Systems Technology}, 26(2): 415--426.

\bibitem{Squire2020}
M. D. Squire, et al. 2020. Cyclomatic Complexity and Basis Path Testing Study. \textit{NASA, NASA/TM-20205011566}: 1--27.

\bibitem{DO178C}
RTCA. DO-178C Software Considerations in Airborne Systems and Equipment Certification. RTCA.

\bibitem{ISO26262}
ISO. ISO 26262 Road Vehicles - Functional Safety. ISO.

\bibitem{IEC61508}
IEC. IEC 61508 Functional Safety of Electrical/Electronic/Programmable Electronic Safety-Related Systems. IEC.

\bibitem{IEC62304}
IEC. IEC 62304 Medical Device Software - Software Life Cycle Processes. IEC.

\bibitem{EN50128}
CENELEC. EN 50128 Railway Applications - Communication, Signaling and Processing Systems - Software for Railway Control and Protection Systems. CENELEC.

\bibitem{Chilenski2001}
J. J. Chilenski. 2001. An Investigation of Three Forms of the Modified Condition Decision Coverage (MCDC) Criterion. \textit{DOT/FAA, AR-01(18)}: 1--214.

\bibitem{CAST2001}
CAST. 2001. Rationale for Accepting Masking MC/DC in Certification Projects. \textit{CAST, CAST-6}: 1--9.

\bibitem{Rajan2014}
A. Rajan, M. Staats, G. Gay, M. Whalen, and M. P. E. Heimdahl. 2014. The Effect of Program and Model Structure on the Effectiveness of MC/DC Test Adequacy Coverage. \textit{ACM TOSEM}, 25(3): 1--34.

\bibitem{Hayhurst2001}
K. J. Hayhurst, D. S. Veerhusen, J. J. Chilenski, and L. K. Rierson. 2001. A Practical Tutorial on Modified Condition/Decision Coverage (MC/DC). \textit{NASA Technical Memorandum}, 210876: 1--78.

\bibitem{CAST2002}
CAST. 2002. What is a "decision" in application of modified condition/ decision coverage and decision coverage (dc). \textit{CAST, CAST-10}: 1--8.

\bibitem{Cao2024}
C. J. Cao, H. H. Huang, F. Liu, Q. Zhang, and Z. Hao. 2024. Multi-task collaborative method based on manifold optimization for automated test case generation based on path coverage. \textit{Expert Systems with Applications}, 251(123932): 1--10.

\bibitem{Shekhawat2021}
S. Shekhawat, A. Iqbal, U. Srinivsan, and P. Menon. 2021. Automation of MC/DC Coverage Test Case Suite Deploying the Optimal Strategic Choice for Tool Development. In \textit{ICT with Intelligent Applications. Smart Innovation, Systems and Technologies}, 248: 433--443.

\bibitem{Barisal2021}
S. K. Barisal, A. Dutta, S. Godboley, B. Sahoo, and D. P. Mohapatra. 2021. MC/DC guided Test Sequence Prioritization using Firefly Algorithm. \textit{Evolutionary Intelligence}, 14: 105--118.

\bibitem{Godboley2017}
S. Godboley, A. Dutta, D. P. Mohapatra, and R. Mall. 2017. J3 Model: A Novel Framework for Improved Modified Condition/Decision Coverage Analysis. \textit{Computer Standards \& Interfaces}, 50: 1--17.

\bibitem{Kangoye2015}
S. Kangoye, A. Todoskoff, M. Barreau, and P. Germanicus. 2015. MC/DC Test Case Generation Approaches for Decisions. In \textit{ACM ASWEC 2015 24th Australasian Software Engineering Conference}, 74--80.

\bibitem{Haque2014}
A. Haque, I. Khalil, and K. Z. Zamli. 2014. An Automated Tool for MC/DC Test Data Generation. In \textit{ISCI}, 130--135.

\bibitem{Ghani2009}
K. Ghani and J. A. Clark. 2009. Automatic Test Data Generation for Multiple Condition and MCDC Coverage. In \textit{IEEE ICSEA}, 152--157.

\bibitem{Awedikian2009}
Z. Awedikian, K. Ayari, and G. Antoniol. 2009. MC/DC Automatic Test Input Data Generation. In \textit{GECCO}, 1657--1664.

\bibitem{Kitamura2018}
T. Kitamura, Q. Maissonneuve, E. H. Choi, C. Artho, and A. Gargantini. 2018. Optimal Test Suite Generation for Modified Condition Decision Coverage Using SAT Solving. In \textit{SAFECOMP}, 11093: 123--138.

\bibitem{Yang2018}
L. Yang, J. Yan, and J. Zhang. 2018. Generating Minimal Test Set Satisfying MC/DC Criterion via SAT Based Approach. In \textit{ACM SAC}, 1899--1906.

\bibitem{Jaffar2019}
J. Jaffar, S. Godboley, and R. Maghareh. 2019. Optimal MC/DC Test Case Generation. In \textit{IEEE/ACM 41st International Conference on Software Engineering: Companion Proceedings (ICSE-Companion)}, 288--289.

\bibitem{Golla2024}
M. R. Golla and S. Godboley. 2024. Automated SC-MCC Test Case Generation using Bounded Model Checking for Safety-Critical Applications. \textit{Expert Systems with Applications}, 238: 122033--122052.

\bibitem{Golla2025}
M. R. Golla, A. Das, S. Godboley, and P. R. Krishna. 2025. Poster: Reporting Unique-Cause MC/DC Score using Formal Verification. In \textit{2025 IEEE Conference on Software Testing, Verification and Validation (ICST)}.

\bibitem{Ahishakiye2021}
F. Ahishakiye, J. I. Requeno Jarabo, L. M. Kristensen, and V. Stolz. 2021. MC/DC Test Cases Generation Based on BDDs. In \textit{Dependable Software Engineering. Theories, Tools, and Applications}, 13071: 178--197.

\bibitem{Kaur2011}
A. Kaur and S. Goyal. 2011. A genetic algorithm for regression test case prioritization using code coverage. \textit{International Journal on Computer Science and Engineering}, 3(5): 1839--1847.

\bibitem{Cegin2020}
J. \v{C}egi\v{n} and K. Rasto\v{c}n\'{y}. 2020. Test Data Generation for MC/DC Criterion using Reinforcement Learning. In \textit{IEEE ICSTW}, 354--357.

\bibitem{Sartaj2025}
H. Sartaj, M. Z. Iqbal, A. A. A. Jilani, and M. U. Khan. 2025. Search-Based MC/DC Test Data Generation With OCL Constraints. \textit{Software Testing, Verification and Reliability}, 35: 21906.

\bibitem{Barisal2024}
S. K. Barisal, A. Dutta, S. Godboley, B. Sahoo, and D. P. Mohapatra. 2024. SMUP: A technique to improve MC/DC using specified patterns. \textit{Computers and Electrical Engineering}, 120: 109706.

\bibitem{Weyuker1994}
E. Weyuker, T. Goradia, and A. Singh. 1994. Automatically Generating Test Data from a Boolean Specification. \textit{IEEE Transactions on Software Engineering}, 20(5): 353--363.

\bibitem{Santhanam2007}
V. Santhanam, J. J. Chilenski, R. Waldrop, T. Leavitt, and K. J. Hayhurst. 2007. SOFTWARE VERIFICATION TOOLS ASSESSMENT STUDY. \textit{DOT/FAA, AR-06(54)}: 1--139.

\bibitem{DO330}
RTCA. DO-330 Software tool qualification considerations. RTCA.

\bibitem{Chen2020}
Z. Chen, H. Washizaki, and Y. Fukazawa. 2020. Automated Tool for Revising Masking MC/DC Test Suite. In \textit{ISSREW}, 157--158.

\bibitem{Godboley2021a}
S. Godboley, J. Jaffar, R. Maghareh, and A. Dutta. 2021. Toward Optimal MC/DC Test Case Generation. In \textit{ISSTA}, 505--516.

\end{thebibliography}

\clearpage
\appendix
\section{Full SBE Benchmark Expressions}
\label{app:sbe}

\lstset{
  basicstyle=\ttfamily\scriptsize,
  breaklines=true,
  breakatwhitespace=true,
  columns=fullflexible,
  keepspaces=true,
  aboveskip=2pt,
  belowskip=2pt,
  xleftmargin=0pt,
  frame=single 
}

\begin{lstlisting}[caption={SBE No. 1 (TCAS-II-derived)}, label={lst:sbe1}]
!(a && b) && (c && !d && !e || !f && g && !h || !i && !j && !k) && (l && m && (n || o) && p || q && (r || s) && !t || u && (v || w))
\end{lstlisting}

\begin{lstlisting}[caption={SBE No. 2 (TCAS-II-derived)}, label={lst:sbe2}]
a && (!b || !c) && d || e
\end{lstlisting}

\begin{lstlisting}[caption={SBE No. 3 (TCAS-II-derived)}, label={lst:sbe3}]
a && (!b || !c || d && e && !(!f && g && h && !i || !j && k && l) && !(!m && n && o && p || !q && !r && s)) || t
\end{lstlisting}

\begin{lstlisting}[caption={SBE No. 4 (TCAS-II-derived)}, label={lst:sbe4}]
(!a && b || c && !d) && !(e && f) && !(g && h) && !(i && j) && ((k && l || m && n) && o && (!p || !q && !r || !s && (!t || !u)))
\end{lstlisting}

\begin{lstlisting}[caption={SBE No. 5 (TCAS-II-derived)}, label={lst:sbe5}]
(!a && b || c && !d) && !(e && f) && !(g && h) && ((i && j || k && l) && m && (n && o || !p && q))
\end{lstlisting}

\begin{lstlisting}[caption={SBE No. 6 (TCAS-II-derived)}, label={lst:sbe6}]
!(a && b) && (!c && d && !e && !f && (g && h || !i && j))
\end{lstlisting}

\begin{lstlisting}[caption={SBE No. 7 (TCAS-II-derived)}, label={lst:sbe7}]
a && !b && !c && d && !e && f && (g || !h && (i || j)) && !(k && l || !m && n || o)
\end{lstlisting}

\begin{lstlisting}[caption={SBE No. 8 (TCAS-II-derived)}, label={lst:sbe8}]
a && !b && !c && !((d && (e || !f && (g || h))) || i && (j || !k && (l || m)) && !n && !o) && !(p && q || !r && s && !t)
\end{lstlisting}

\begin{lstlisting}[caption={SBE No. 9 (TCAS-II-derived)}, label={lst:sbe9}]
a && !b && !c && (d && (e || !f && (g || h)) && (!i && !j || k) || !l) && (m && n || !o && p && !q)
\end{lstlisting}

\begin{lstlisting}[caption={SBE No. 10 (TCAS-II-derived)}, label={lst:sbe10}]
a || b || c || !d && !e && f && g && !h && !i || j && (k || l) && !m
\end{lstlisting}

\begin{lstlisting}[caption={SBE No. 11 (TCAS-II-derived)}, label={lst:sbe11}]
a && b && (c || d) && e || f && (g || h) && !i || j && (k || l)
\end{lstlisting}

\begin{lstlisting}[caption={SBE No. 12 (TCAS-II-derived)}, label={lst:sbe12}]
a && ((b || c || d) && e || f && g || h && (i || j || k || l)) || (m || n) && (o || p || q) && r
\end{lstlisting}

\begin{lstlisting}[caption={SBE No. 13 (TCAS-II-derived)}, label={lst:sbe13}]
(a && b || c && d) && e && (f || (g && (h && i || j && k)))
\end{lstlisting}

\begin{lstlisting}[caption={SBE No. 14 (TCAS-II-derived)}, label={lst:sbe14}]
(a && b || c && d) && e && (f && g || !h && i)
\end{lstlisting}

\begin{lstlisting}[caption={SBE No. 15 (TCAS-II-derived)}, label={lst:sbe15}]
!a && b && !c && !d && (e && f || !g && h)
\end{lstlisting}

\begin{lstlisting}[caption={SBE No. 16 (Randomly generated)}, label={lst:sbe16}]
(((a || b) || c && !(!(d && e)) && f || g) || !h || i || !((j || !(!((k && (l && !m))) || n && o || p || q && r || s))))
\end{lstlisting}

\begin{lstlisting}[caption={SBE No. 17 (Randomly generated)}, label={lst:sbe17}]
((a && b && c && d) && !(e || f) || (!(!g) && !h && i) && j && k)
\end{lstlisting}

\begin{lstlisting}[caption={SBE No. 18 (Randomly generated)}, label={lst:sbe18}]
!(a || b) || (c && d && !e || !f && g || (!h || i) && j && !k) || (l && m && (n || o) && p || q && (r || s) && !t && u && (v && w) || x)
\end{lstlisting}

\begin{lstlisting}[caption={SBE No. 19 (Randomly generated)}, label={lst:sbe19}]
!(a || (b && !c)) && (!(!(d && e)) && f || g || (!((h || i) || !j) && !(!k))) || !(!(l && m)) && n && o && !p || (!q || r)
\end{lstlisting}

\begin{lstlisting}[caption={SBE No. 20 (Randomly generated)}, label={lst:sbe20}]
((!((a && b)) && !((c && (!(!d) || e && f)) && ((!g && (!h || !(!i) || (j || k))) && !l || m) && n || !(!o) || p)) || !(!((q || r || s) || !t && !((!(!u) || v || w)) || !(!x) && !y && z)))
\end{lstlisting}

\begin{lstlisting}[caption={SBE No. 21 (Randomly generated)}, label={lst:sbe21}]
(a || !b || !(c && d) || (e && f) || g) && (h && i && j) || (k && l && m)
\end{lstlisting}

\begin{lstlisting}[caption={SBE No. 22 (Randomly generated)}, label={lst:sbe22}]
a || b && c || d && e || (!(!(f || g && h) || i && !(!j) && !k) && l || m || !n && o && p && q)
\end{lstlisting}

\begin{lstlisting}[caption={SBE No. 23 (Randomly generated)}, label={lst:sbe23}]
(!a && b || c) || d && e || !f && g
\end{lstlisting}

\begin{lstlisting}[caption={SBE No. 24 (Randomly generated)}, label={lst:sbe24}]
a && !b && c || d && !e && !((f && g || (!h && i)) && j || k)
\end{lstlisting}

\begin{lstlisting}[caption={SBE No. 25 (Randomly generated)}, label={lst:sbe25}]
!((a && b)) && c || (!(((d || e) || f) && g) && h && !(!(i && j))) || k && l || m || n || o && (p && q && r)
\end{lstlisting}

\end{document}